\documentclass[]{spie}  

\usepackage{amsmath,amsfonts,amssymb}
\usepackage{graphicx}
\usepackage[colorlinks=true, allcolors=blue]{hyperref}
\usepackage{caption}
\usepackage{subcaption}
\usepackage{bibstyle}
\usepackage[bottom]{footmisc}

\title{Intensity Interferometry at Calern and beyond:\\ progress report}

\author[a]{Nolan Matthews}
\author[b]{Jean-Pierre Rivet}
\author[a]{Mathilde Hugbart}
\author[a]{Guillaume Labeyrie}
\author[a]{Robin Kaiser}
\author[b]{Olivier Lai}
\author[b]{Farrokh Vakili}
\author[c]{David Vernet}
\author[d]{Julien Chab\'{e}}
\author[d]{Cl\'{e}ment Courde}
\author[e]{Nicolas Schuhler}
\author[e]{Pierre Bourget}
\author[a]{William Guerin}
\affil[a]{Universit\'{e} C\^{o}te d'Azur, CNRS, Institut de Physique de France, France}
\affil[b]{Universit\'{e} C\^{o}te d'Azur, Observatoire de la C\^{o}te d'Azur, CNRS, Laboratoire Lagrange, France}
\affil[c]{Universit\'{e} C\^{o}te d'Azur, Observatoire de la C\^{o}te d'Azur, CNRS, UMS Galil\'{e}e, France}
\affil[d]{Universit\'{e} C\^{o}te d'Azur, Observatoire de la C\^{o}te d'Azur, CNRS, Laboratoire G\'{e}oazur, France}
\affil[e]{European Southern Observatory, Vitacura, Chile}

\authorinfo{\hspace{-18pt}Send correspondence to: \\ N.M.: nolankmatthews@gmail.com //  W.G.: william.guerin@inphyni.cnrs.fr \vspace{10pt}}

\begin{document} 
\maketitle

\begin{abstract}
We present the current status of the I2C stellar intensity interferometer used towards high angular resolution observations of stars in visible wavelengths. In these proceedings, we present recent technical improvements to the instrument, and share results from ongoing campaigns using arrays of small diameter optical telescopes. A tip-tilt adaptive optics unit was integrated into the optical system to stabilize light injection into an optical fiber. The setup was successfully tested with several facilities on the Calern Plateau site of the Observatoire de la C\^{o}te d'Azur. These include one of the 1\,m diameter telescopes of the C2PU observatory, a portable 1\,m diameter telescope, and also the 1.5\,m MéO telescope. To better constrain on-sky measurements, the spectral transmission of instrument was characterized in the laboratory using a high resolution spectrograph. The system was also tested with two of the auxiliary telescopes of the VLTI resulting in successful temporal and spatial correlation measurements of three stars.
\end{abstract}


\section{INTRODUCTION}
\label{sec:intro}  

Intensity interferometry was originally pioneered by Robert Hanbury Brown and Richard Q. Twiss (HBT) in the 1950s, leading to the successful implementation and operations of the Narrabri Stellar Intensity Interferometer that measured the angular diameters of 32 stars \cite{HBT1974}. The technique was subsequently abandoned in astronomy in favor of the far more sensitive techniques that had emerged in stellar amplitude interferometry with separated telescopes\cite{Labeyrie1975} of which has steadily evolved into the current exceptional interferometric facilities present worldwide such as the CHARA\footnote{https://www.chara.gsu.edu/} array and the VLTI\footnote{https://www.eso.org/sci/facilities/paranal/telescopes/vlti.html}. However, over the last decade and a half there has been strong experimental efforts in astronomical HBT interferometry, that can be broadly divided into two complementary approaches. One is to use imaging atmospheric Cherenkov telescopes (IACTs), principally designed for ground based gamma-ray astronomy for intensity interferometry by taking advantage of their large diameters ($>$ 10\,m), and array based layouts\cite{LeBohec2006}. Successful on-sky measurements have been demonstrated with both the MAGIC\cite{Acciari2019} and VERITAS\cite{Abeysekara2020} observatories with key implications for a possible implementation on the future Cherenkov Telescope Array that holds significant promise if operated additionally as an HBT observatory \cite{Dravins2013}. An alternative approach to implementing HBT interferometry is to use optical telescopes whose significantly better optical quality allows the usage of advanced technologies such as state of the art single photon counting detectors. Our I2C consortium\footnote{ https://inphyni.univ-cotedazur.eu/sites/cold-atoms/research/i2c}, primarily consisting of a collaboration of physicists and astronomers in the network of the Université Côte d’Azur, is focused on this latter approach, demonstrating leadership through observations with the facilities on the Calern Plateau site of the Observatoire de la Côte d’Azur.

Our group has made steady progress in the development of an on-sky HBT instrument demonstrating both technical and scientific achievements. Single telescope temporal intensity interferometry measurements were first successfully performed with a 1\,m diameter telescope\cite{Guerin2017}, and soon thereafter extended to spatial intensity interferometry, measuring the HBT correlations between a pair of 1\,m telescopes with a 15\,m baseline at the C2PU observatory\cite{Guerin2018}. Subsequently, the angular diameter of the bright luminous blue variable (LBV) star P Cyg was measured in its H-alpha emission, which provided a distance measurement when combined with a linear diameter obtained from spectroscopic modeling\cite{Rivet2020}. Using an improved version of our instrument, this technique was reapplied to P Cyg and additionally to $\beta$\,Ori.\cite{deAlmeida2022} The long term goal of our group is to continue the progress made thus far towards an instrument which can perform interferometric measurements in visible wavelengths (B, V, R, I photometric bands) with sensitivity to stars brighter than at least magnitude 8 if installed onto major facilities. 

The focus of this proceeding is to provide an update on recent technical improvements made to extend the I2C instrument to additional optical telescopes and to increase the sensitivity to fainter stars, or equivalently with improved precision.  A first order tip-tilt adaptive optics unit was integrated into the optical system to stabilize and improve the overall light collection efficiency of the system. The setup was sequentially tested on several of the telescopes available on the Calern Plateau and allows for coordinated campaigns between the different telescopes, potentially providing up to 4 telescope intensity interferometry observations with semi-configurable baselines ranging approximately from 15 to 150 meters at visible wavelengths. Furthermore, the temporal and spectral characteristics of the instrument were extensively measured in the laboratory. Laboratory HBT correlation peaks were measured at the level of 1\%, thus demonstrating the potential of squared visibility measurements with a precision of 10$^{-2}$. Finally, the system was successfully tested at the VLTI using two of the auxiliary telescopes in observations of three stars. 

\subsection{Brief Theory of Intensity Interferometry}
An intensity interferometer performs the correlation of starlight intensity fluctuations between two telescopes in order to measure the squared visibility. For two telescopes with a projected baseline $r$ between them, the second order coherence function is
\begin{equation}
    g^{(2)} (r,\tau) = \frac{\langle I_1(t) I_2(r,t+\tau)  \rangle}{\langle I_1 \rangle \langle I_2 \rangle}
\end{equation}
where $I_1$ and $I_2$ are time-dependent intensities recorded at the two telescopes with a time-lag $\tau$ between them, and the brackets indicate averaging over time $t$. The second-order coherence function ``\,$g^{(2)}$\,'' is related to the first order coherence function $g^{(1)}$ through the Siegert relation\cite{Siegert1943,Ferreira2020} by
\begin{equation}
    g^{(2)}(r,\tau) = 1 + |g^{(1)}(r,\tau)|^2
\end{equation}
where the first-order coherence function can be separated into spatial and temporal components
\begin{equation}
    g^{(1)}(r,\tau)=V(r) g^{(1)} (\tau)
\end{equation} 
where $V(r)$ is the interferometric visibility of the source, given by the Fourier transform of the source sky brightness distribution. In the case of an unresolved source $V(r)=1$, and the measured second order coherence function will depend only on the temporal component $g^{(1)}(\tau)$ given by the Fourier transform of the measured light spectral density by the Wiener-Khintchine theorem. For linearly polarized thermal light at zero optical path delay $g^{(1)}(\tau=0) = 1$ whereas at large time-lags much greater than the coherence or correlation time $g^{(1)} = 0$ such that there is a ``bunching peak" centered about zero optical path delay with an effective temporal width given by the coherence time. The amplitude of this peak at zero time-lag measures the squared visibility under the condition that the instrumental resolving time is shorter than the coherence time. The coherence time as defined by Mandel \& Wolf (1995)\cite{Mandel1995} can be described by the integral of the squared first-order coherence function
\begin{equation}
    \label{eqn:coh_time}
    T_c = \int |g^{(1)}(\tau)|^2 d\tau = \int |s(\nu)|^2 d\nu
\end{equation}
which by Parseval's theorem is equal to the integral of the squared normalized spectral density $s(\nu)$. For visible light with a bandpass of $\Delta \lambda \sim 1\,$nm the corresponding coherence time is of order 1\,ps, much shorter than what can be achieved with conventional detectors. In this case, a measurement averages over many coherence times and reduces the value of the $g^{(2)}$ peak amplitude at $\tau = 0$ by a factor of $\sim T_c / T_d$ where $T_d$ is the effective time-resolution of the detector. The result is that amplitude of the $g^{(2)}$ peak does not directly measure the squared visibility, but instead the squared visibility times this loss of contrast. The squared visibility can be extracted by dividing the value of the $g^{(2)}(r)-1$ peak measured between telescopes to the $g^{(2)}-1$ peak measured at zero-baseline
\begin{equation}
    |V(r)|^2 = \frac{g^{(2)}(r,\tau=0)-1}{g^{(2)}(r=0,\tau=0)-1}
\end{equation}
under the assumption that the profile of the $g^{(2)}(\tau)$ peak is constant. In practice, we measure the ratio of the area of the $g^{(2)}(r)-1$ peak, to the area of the $g^{(2)}(r=0)-1$ peak for the squared visibility, where at zero-baseline this area should correspond to the coherence time that can be calculated from the detected spectrum as given by Equation~\ref{eqn:coh_time}.

\begin{figure}[b]
    \begin{center}
    \begin{tabular}{c} 
   \includegraphics[width=0.8\linewidth]{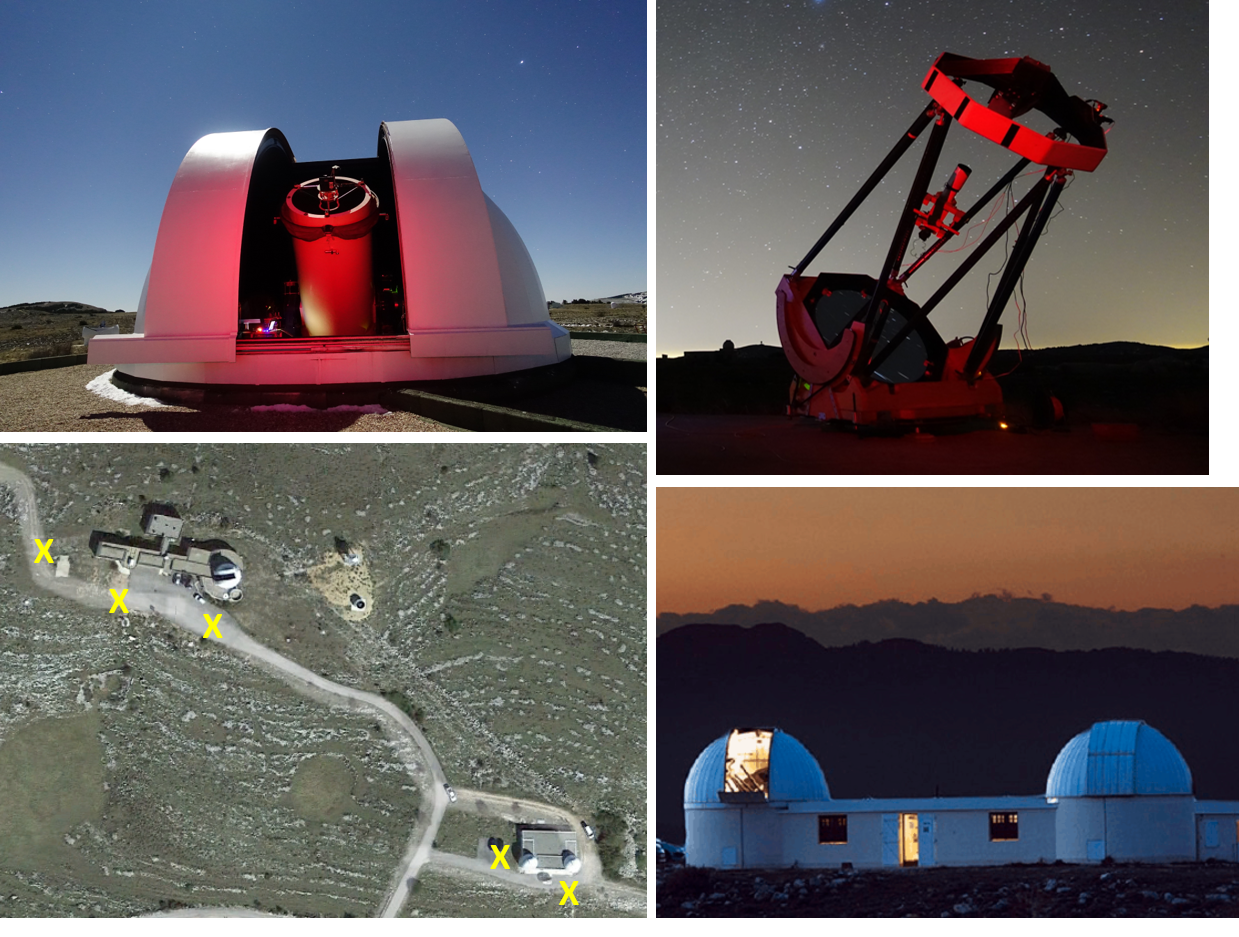}
   \end{tabular}
   \end{center}
   \caption[example] 
   {\label{fig:calernfacilities} Photographs of facilities available for intensity interferometry observations on the Calern Plateau site of the Observatoire de la C\^{o}te d'Azur. These include the M\'{e}O optical metrology station (top-left), the T1M portable telescope (top-right), and the C2PU observatory (bottom right). A satellite photograph (Google) of the facilities are shown in the bottom-left with C2PU in the Southeast, and M\'{e}O at the Northwest. The yellow crosses indicate the locations at which differential GPS positions were obtained, and serve at the current possible positions of the T1M telescope.}
\end{figure}

\section{Intensity Interferometry on the Calern Plateau}

Observations with the I2C instrument are primarily conducted using the facilities of the Observatoire de la C\^{o}te d'Azur available on the Calern Plateau site located in southern France. Previously reported observations were performed with the Centre P\'{e}dagogique Plan\`{e}te Univers (C2PU) facility consisting of a pair of effectively identical 1\,m diameter optical telescopes that are separated by a fixed 15\,m separation on a nearly East-West baseline\footnote{https://www.oca.eu/fr/c2pu-accueil}. Recently, technical intensity interferometry tests were successfully performed with the 1.5\,m diameter M\'{e}O (M\'{e}trologie Optique) telescope that is located approximately 150\,m to the North-West of the Omicron (westernmost) telescope of C2PU. Similar tests were performed with the T1M telescope, a portable 1\,m diameter Newtonian telescope on a Dobson-type fully motorized azimuthal mount that enables configurable baselines to extend the accessible sampling of the uv-plane. These facilities are shown in Figure~\ref{fig:calernfacilities}. To help aid in the planning of future observations we have been able to incorporate a custom array configuration file within the interferometric observational planning tool ASPRO\cite{Bourges2016} to coordinate the optimal baseline configurations for a given target. 

\subsection{Experimental setup}
\label{sec:exp_setup}

\begin{figure}[b]
\centering
\begin{subfigure}{.65\linewidth}
  \centering
  \includegraphics[width=\linewidth]{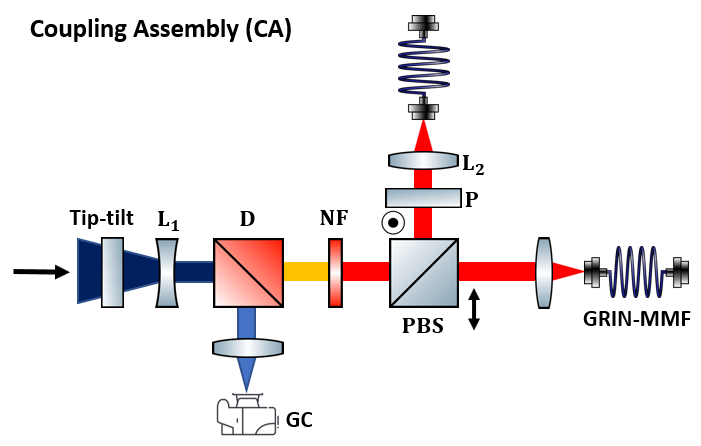}
\end{subfigure}%
\begin{subfigure}{.35\linewidth}
  \centering
  \includegraphics[width=\linewidth]{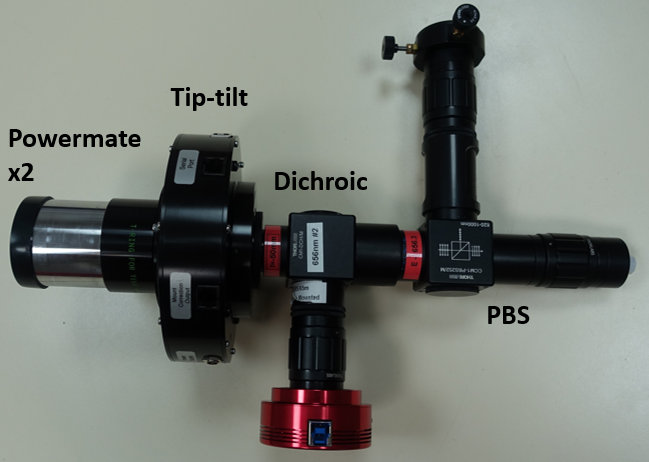}
\end{subfigure}
\caption{\label{fig:coupling_assembly} The left figure sketches the general operation of the coupling assemblies (CAs). The tip-tilt correction is applied to the incident beam to provide corrections in the XY plane. The diverging lens ($L_1$) produces a collimated beam required to properly filter the light with the narrowband filter ($NF$), typically of width $\Delta \lambda \sim 1\,$nm. A dichroic beamsplitter ($D$) reflects blue light towards a guiding camera used for the tip-tilt control. The transmitted red light is passed through the narrowband filter, and then is separated into two orthogonal linear polarization states by a polarizing beam splitter ($PBS$). On the reflected beam, an additional linear polarizer is included ($P$) to further reject the complementary polarization state. The light is then coupled to a graded-index multi-mode fiber (GRIN-MMF) with a 100$\,\mu$m core diameter using a converging lens ($L_2$). The right image displays a photograph of the CA for the T1M telescope that adapts to the output port of the telescope. Note that the incident beam does not always come directly from the telescope: The light is first modified by either a Barlow system (shown by the leftmost cylindrical element in the right photograph) or focusing lens, as described in Section~\ref{sec:exp_setup}.} 
\end{figure}

To enable each telescope for intensity interferometry observations we install a coupling assembly (CA) at the focus of each telescope. These CAs differ slightly for each telescope to account for differences in the primary focal length. The general structure of the coupling assembly is pictured in Figure~\ref{fig:coupling_assembly}. The main purpose of the CA is to perform spectral and polarization filtering, and subsequent injection of the filtered starlight into a multimode optical fiber. Each CA separates the light into two orthogonal polarization states which are each injected into an APC optical fiber to allow for either polarization resolved measurements or improve sensitivity by co-adding the $g^{(2)}$ measurements made in each polarization mode. With exception to the C2PU telescopes, the input converging beam shown in Figure~\ref{fig:coupling_assembly} does not come directly from the telescopes. On the T1M telescope ($F=3\,$m), a corrected Barlow device (Powermate x2) is inserted upstream from the CA, to enhance the focal length. On the MéO telescope ($F=31\,$m), the focal length is first reduced by a converging lens placed upstream from the CA. Furthermore, the focal length of the collimating lens (optic $L_1$ in Figure~\ref{fig:coupling_assembly}) is different on both telescopes ($f=-50$\,mm for the T1M and C2PU, and $f=-75\,$mm on MéO). 

The fiber for both polarization states are fed to 50/50 fiber splitters that allows a zero-baseline $g^{(2)}$ measurement to be compared against expectation from the spectrum. The outputs of the fiber splitters are connected to single photon avalanche diode (SPAD) detectors (Excelitas SPCM-ARQH) with a characteristic timing resolution of $\sim 500\,$ps. The detection of a photon generates an electronic pulse that is then fed to a time to digital converter (Swabian Instruments) that records the pulse arrival time and performs the cross-correlation between all pairs of detectors measuring parallel polarization states. The $g^{(2)}$ correlations are recorded to disk typically every 10 seconds over a time-lag window of $\pm 300\,$ns with 10\,ps binning. A software analysis then shifts each correlation function in time by the instrumental and geometrical optical path delays such that they are centered about zero optical path delay. The time-corrected correlations are then averaged together in order to resolve the $g^{(2)}$ bunching peak. 

\subsection{Inclusion of tip-tilt adaptive optics}

In our setup, the size of the multimode fiber core corresponds to an on-sky angle of $\sim 7\,"$. Any imperfection in the telescope tracking results in a loss of collected flux, degrading the overall sensitivity. The tracking precision of each telescope varies, but is especially pronounced in the case of the T1M due to a non-permanent mounting that allows for portability but makes it especially subject to environmental conditions, particularly wind. In past observations, any small imperfections in the tracking were corrected using manual commands sent to the telescope that were typically required every few minutes. 

To remove the need for manual corrections, and to improve the overall light collection efficiency, a first-order tip-tilt adaptive optics unit (Starlight Xpress AO-USB) was integrated into the CA and can be seen in Figure~\ref{fig:coupling_assembly}. The tip-tilt optic consists of a 60\,mm diameter glass plate with a thickness of 13\,mm that can be tilted by small stepper motors in order to displace the optical beam in the XY plane. The range of displacements in the XY plane is $\pm\,0.15$\,mm or equivalently 1.5 times larger than the MMF core diameter. If a larger correction is needed, a command can be sent from the ST4 port output of the tip-tilt module to adjust the telescope position in order to bring the image of the star back within the range of accessible displacements provided by the tip-tilt. 

The tip-tilt is controlled via a program written in Python that applies commands based on the position of the star image in the CA guiding camera relative to a desired setpoint. Both the guiding camera and the tip-tilt are connected via USB to a laptop that runs the tip-tilt control software. The tip-tilt can receive commands up to 5\,ms per increment limiting the response to frequencies less than 200 Hz. In practice, we are able to achieve a rate correction of $\sim40\,$Hz which takes into account the image processing time. This rate is more than sufficient to correct for the slow imperfections in the telescope tracking, but cannot account for faster atmospheric effects. 

\begin{figure}[t]
\centering
\begin{subfigure}{.52\linewidth}
  \centering
  \includegraphics[width=\linewidth]{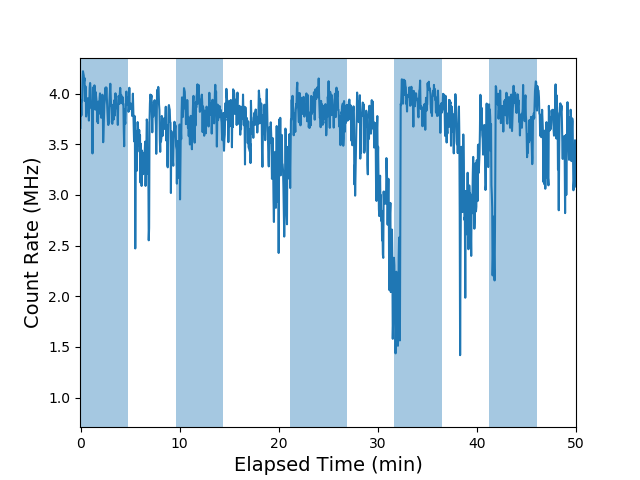}
  \label{fig:sub1}
\end{subfigure}%
\begin{subfigure}{.52\linewidth}
  \centering
  \includegraphics[width=\linewidth]{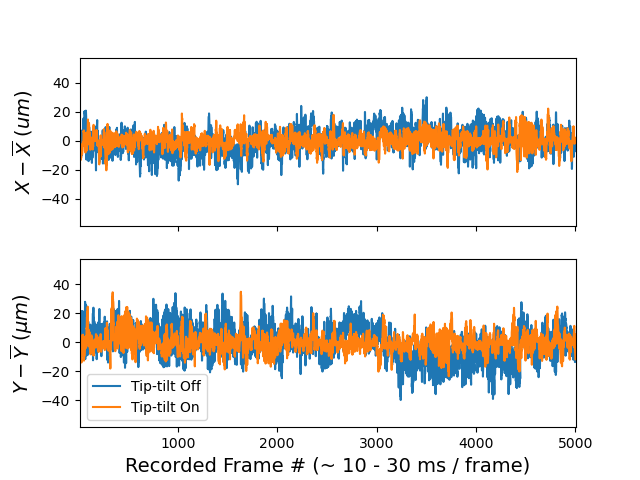}
  \label{fig:sub2}
\end{subfigure}
\caption{\label{fig:tiptilt_onoff} The left figure shows the detected count rate in units of photons per second as a function of time. The regions shaded in blue/white indicate when the tip-tilt control was active/deactive, respectively. The right figure shows the displacement of the star centroid about the mean along the X and Y axes over a duration of 1 minute for when the tip-tilt control was applied (orange) and deactivated (blue).}
\end{figure}

Initial tests of the tip-tilt were performed in late Summer 2021 / Fall 2021 with several of the telescopes available at Calern. To quantify the improvement in light collection efficiency we recorded data, taken using the Omicron telescope of the C2PU, where the tip-tilt control was alternated on and off approximately every 5 minutes for a duration of one hour. The main results of these tests are shown in Figure~\ref{fig:tiptilt_onoff}. We find a modest improvement of 8\% in the overall efficiency when comparing the mean count rate during the on cycles to the off cycles. In addition, the average displacement of the star image centroid on the guiding camera was measured over a duration of approximately 1 minute. The tip-tilt reduced the root mean square radial fluctuations about the mean from 15.0\,$\mu$m to 8.4\,$\mu$m, or equivalently from 15\,\% to 8.4\,\% of the fiber core diameter. However, we note that the amount of improvement in fiber injection efficiency is strongly dependent on additional factors, and in particular wind and seeing conditions which affect each telescope differently to varying degrees. Nevertheless, the inclusion of the tip-tilt provided stable light injection into the fiber for periods of several hours without any manual correction, thus achieving the primary goal. 

\begin{figure}[t]
    \centering
    \includegraphics[width=0.8\linewidth]{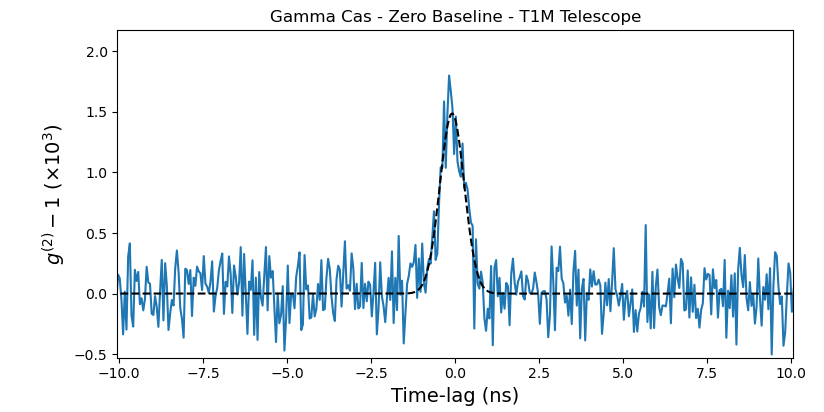}
    \caption{Temporal $g^{(2)}(\tau)$ measurement of Gam Cas observed by the portable T1M telescope. The data (solid blue line) are fit by a Gaussian (dotted black line).}
    \label{fig:gamcas_t1m}
\end{figure}

During January 2022 an observational campaign was performed on $\gamma\,$Cas in it's H$\alpha$ emission line using the M\'{e}O and T1M telescopes. The source was observed in multiple baseline configurations over a duration of approximately 78 hours of observation. A clear $g^{(2)}$ temporal bunching peak was observed with both telescopes. Here, we show the result for the T1M telescope in Figure~\ref{fig:gamcas_t1m}. A full analysis of the data, including reduction of the spatial correlations between the two telescopes with subsequent modeling, is still pending and will be reported in the future. However, the zero-baseline $g^{(2)}$ peak reported here demonstrates the usefulness of the tip-tilt optic to enable intensity interferometry observations onto additional telescopes.   

\section{Laboratory Developments}

In parallel to the observational campaigns, there have been several developments made in the laboratory to provide a robust characterization of the instrument and examine any limiting sensitivity factors. In particular, a high resolution spectrograph was introduced into our laboratory setup to experimentally characterize the spectral transmission of the narrowband interferometric filters used in our CA. In addition, we perform high precision measurements of the $g^{(2)}$ bunching peak at the level of 1-2\%.     

\subsection{Spectral characterization of narrow-band filters}
\label{sec:spectra}

Squared visibilities are extracted by the ratio of the area of the $g^{(2)}$ peak measured between telescopes to that of a zero-baseline correlation. The area of the $g^{(2)}$ peak at zero-baseline is in principle given by the coherence time of the light that is dependent on the detected optical spectrum. To validate the on-sky zero baseline measurement it is therefore useful to properly characterize the spectral transmission of the system. In general, this necessarily includes all instrumental and physical contributions such as the source spectrum and instrumental throughput. The instrumental throughput is primarily given by the transmission of our narrowband interferometric filters, typically of width $\Delta \lambda \sim 1$\,nm. Other instrumental factors such as the mirror reflectivity, detector response, and fiber transmission display weak wavelength dependence over the transmission band of the filter and are assumed as constant. 

To characterize the filter transmission, we constructed a high resolution spectrograph by modifying a monochromator (THR 1500) in which the output beam, normally focused onto the output slit, is instead focused onto a camera. The spectrograph was calibrated using a wavelength tunable laser diode in conjunction with an optical spectrum analyzer. The attained spectral resolution was less than 0.01\,nm at 780\,nm ($R \sim 80000$) which is more than sufficient to characterize our filters. To test the spectral transmission of the filters, we first inject thermal light from a halogen lamp into a multimode fiber. The output beam of the fiber is collimated and directed into our CA. The CA performs the spectral and polarization filtering, and then re-injects the light into a different multi-mode fiber. This fiber is fed to an optics bench at the input of the spectrograph, which conjugates the image of the output fiber and refocuses it onto the input slit of the spectrograph.

\begin{figure}[t]
    \centering
    \includegraphics[width=0.85\linewidth]{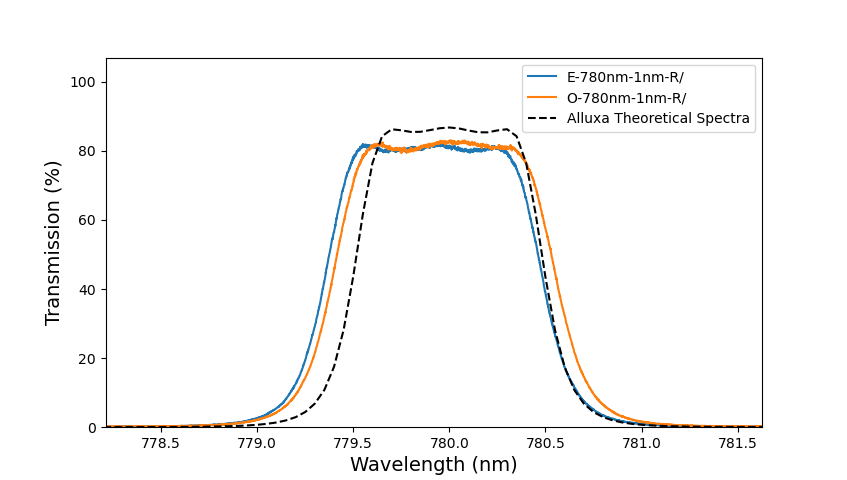}
    \vspace{3pt}
    \caption{\label{fig:spectra_780} Measured spectral transmission of two $\lambda_0 \sim$ 780\,nm filters (in blue and orange) compared with the theoretical spectra (dotted black line).}
\end{figure}

To measure the spectra, we take a series of three images consisting of raw spectra, a flat-field correction, and a dark frame. The flat-field image is performed by recording an image with the narrowband filter removed from the beam path, and the dark image is formed by recording an image with the input beam blocked. The corrected spectrum is then obtained by

\begin{equation}
    T(\lambda) = \frac{\sum_i R(i,\lambda) - \sum_i D(i,\lambda)}{\sum_i F(i,\lambda) - \sum_i D(i,\lambda)}
\end{equation}
where $R,D,F$ are the raw, dark, and flat-field images with spatial coordinate $i$ and wavelength coordinate $\lambda$. An example of the spectral transmission measured in the laboratory are shown in Figure~\ref{fig:spectra_780}. The data correspond to the measurement of two spectral filtering setups, each one used in separate coupling assemblies. Each setup consists of two sequential narrowband filters where both are specified with a central wavelength of 780\,nm, but with different bandpasses of 1\,nm and 10\,nm. The wider bandpass filter is included in order to block out of band transmission, which is pronounced for the 1\,nm bandpass filter. We find that the two filtering setups have a comparable peak transmission of $\sim 82\,$\%. This value is compatible with the vendor specified transmission of $>90\,$\% for a single filter, such that for both filters in a given setup the total transmission should be $>81\,$\%. In Figure~\ref{fig:spectra_780} the theoretical spectra shown assumes 95\,\% transmission in both filters. In addition, the measured spectral transmission shows that the central transmitted wavelength are shifted with respect to the theoretical expectation, and furthermore to one another by approximately 0.1\,nm. We note these measurements are consistent with similar shifts seen in lower spectral resolution measurements from the vendor. If these non overlapping transmission bands are unaccounted for, this will lead to a underestimation of the squared visibility. Similar tests were performed on other filters, including the H$\alpha$ filters used in on-sky observations\cite{Rivet2020,deAlmeida2022}, and were found to be in close agreement with each other.  

\subsection{Precision measurements of HBT peaks}

To fully characterize the experimental setup we perform measurements of artificial stars in the laboratory. The artificial star simulates a point-like thermal source and thus can be used to compare the measured $g^{(2)}$ bunching area to the coherence time expected from the spectral transmission, and against on-sky observations of the $g^{(2)}$ zero-baseline measurement of stellar targets. Furthermore, it allows testing for any systematic effects such as spurious correlations, or detector variability, that may be difficult to characterize during on-sky observations. The artificial star is generated by first injecting light from a halogen lamp into a single mode fiber. The output beam of the fiber is collimated and then directed into our CA in order to measure the $g^{(2)}$ function. We record the $g^{(2)}$ correlation over a wide range of time lags from $\pm$300\,ns with 10\,ps resolution. Correlations are typically recorded every minute over duration a few to a hundred hours. Once the acquisition is complete, we analyze the data by summing all correlations together, and then normalizing in order to obtain $g^{(2)}$. The reduced data is fitted to a Gaussian in order to extract a bunching peak area to measure the coherence time. 

\begin{figure}[t]
\centering
\begin{subfigure}{.52\linewidth}
  \centering
  \includegraphics[width=\linewidth]{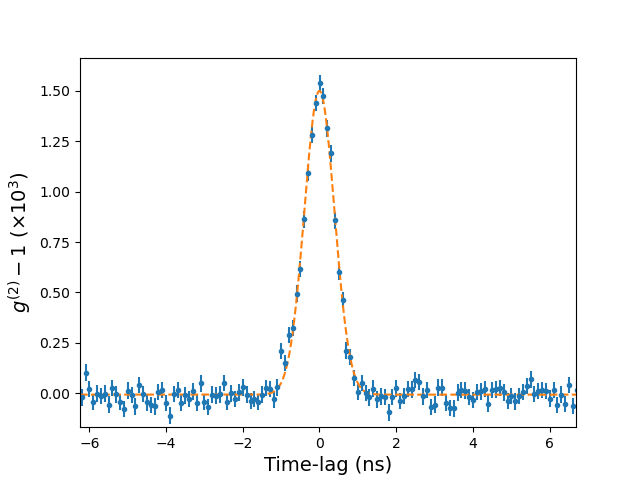}
  \label{fig:sub1}
\end{subfigure}%
\begin{subfigure}{.52\linewidth}
  \centering
  \includegraphics[width=\linewidth]{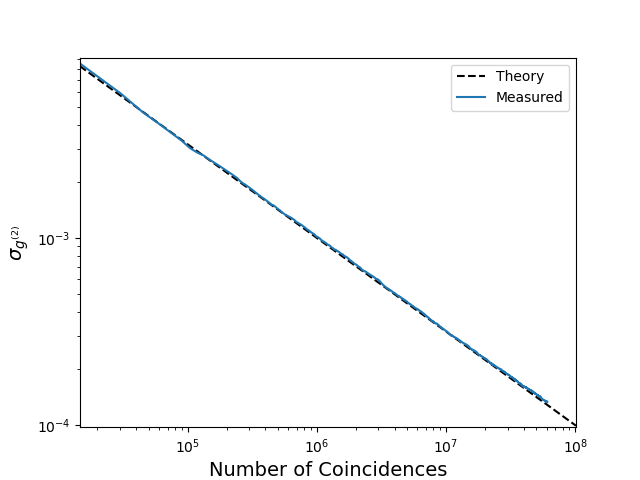}
  \label{fig:sub2}
\end{subfigure}
\caption{\label{fig:g2_lab_measurements} The left figure shows the measured $g^{(2)}$ bunching peak in the blue points as a function of the time-lag, binned into 100\,ps bins. A data is fit by a Gaussian which is shown by the dotted orange line. The right figure shows the measured uncertainty in $g^{(2)}$ as a function of the number of coincidences shown in blue where the dotted black line is the expected uncertainty due to shot-noise fluctuations. The uncertainty is measured by taking the root mean square fluctuations of $g^{(2)}$ over time-lags in which the bunching peak is negligible.}
\end{figure}

Figure~\ref{fig:g2_lab_measurements} shows the measured $g^{(2)}$ bunching peak for one of the $\lambda_0 = 780\,$nm filters described in Section~\ref{sec:spectra}. The data was taken continuously over a duration of 73.9 hours. Figure~\ref{fig:g2_lab_measurements} also shows the measured uncertainty as function of the total number of coincidences. This uncertainty is obtained by taking the root mean square fluctuations of the accumulated $g^{(2)}$ value over time-lags in which the bunching peak is negligible, taken here as all time-lags between $3< |\tau| < 20$\,ns, where $\tau$ is the time-lag difference from the $g^{(2)}$ peak center. The accumulated $g^{(2)}$ peak is fitted by a Gaussian function in which we find a standard deviation of width $392\pm6\,$ps, and an amplitude of $\left(1.504\pm0.017\right) \times 10^{-3}$ corresponding to a bunching area or coherence time of $1.477\pm0.019$\,ps. This value is in very good agreement with the expectation from the measured spectra of 1.474\,ps but notably lower than what is expected from the theoretical spectra of 1.69\,ps. 

The uncertainty in the measured bunching area can be directly corresponded to an uncertainty in squared visibility. Here, we demonstrate in the laboratory a measurement of the bunching area at a level of 1.3\,\%, suggesting that a similar uncertainty can be obtained on stellar sources, provided similar number of photon statistics. While such precision with 1-meter class telescope and single wavelength channel observations under reasonable observations times (e.g. $< 100\,$hours) are so-far limited to bright stars, the measurements are nevertheless robust. The data indicates we are free from systematic effects in our correlation to the level of our measurement precision since the uncertainty is at the theoretical shot-noise limit.  

\section{Technical tests at the VLTI}

\begin{figure}[b]
\centering
\begin{subfigure}{.47\linewidth}
  \centering
  \includegraphics[width=0.78\linewidth]{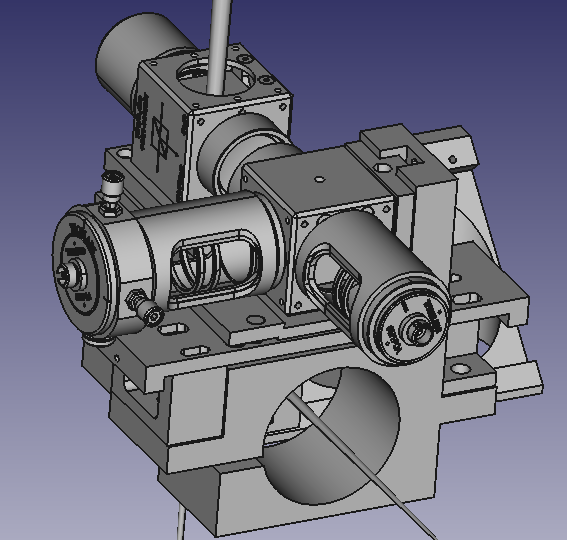}
  \label{fig:sub1}
\end{subfigure}%
\begin{subfigure}{.47\linewidth}
  \centering
  \includegraphics[width=0.9\linewidth]{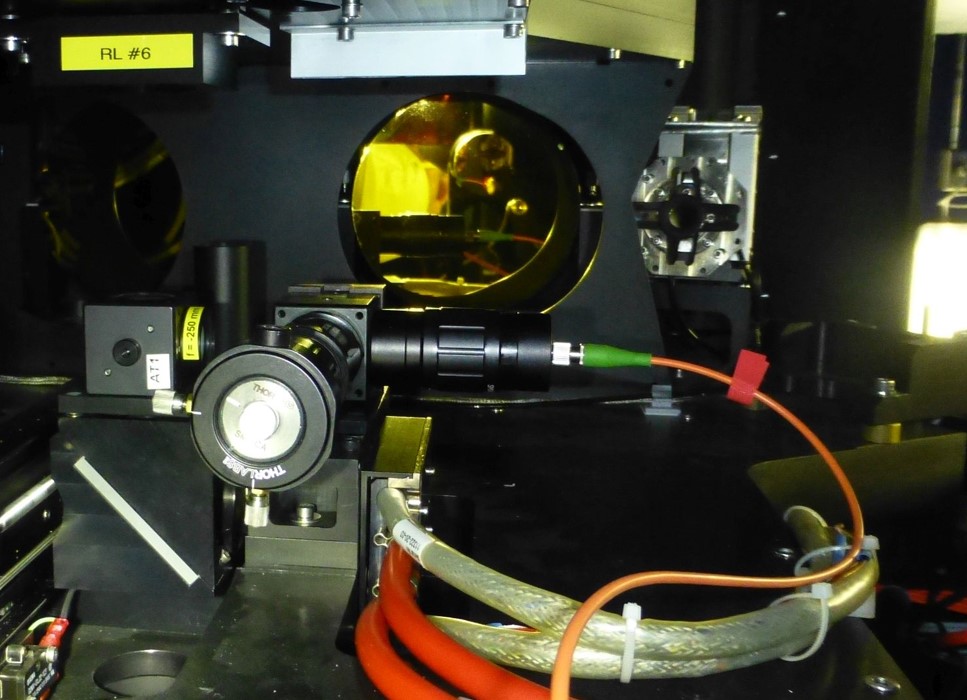}
  \label{fig:sub2}
\end{subfigure}
\vspace{5pt}
\caption{\label{fig:VLT_image} The left photograph displays a CAD drawing of the I2C-VLTI interface + coupling module. The incident beam arrives from the top of the image where it is reflected off a dichroic onto a modified coupling assembly. The coupling assembly performs spectral and polarization filtering of the light, and focuses the light into a multimode fiber that is connected to the detectors. The right photograph shows a picture of the entire interface as mounted in the ROS of one of the ATs.}
\end{figure}

During March 2022 we were given time to install and perform technical observations with our I2C instrument using two of the 1.8\,m diameter auxilary telescopes (ATs) of the Very Large Telescope Interferometer while they were located in their maintenance stations. The main purpose of this mission was to establish the technical feasibility of performing intensity interferometry observations with the ATs. These tests occurred during time in which the ATs were not scheduled for other science programs due to the delay lines being in use with the unit telescopes (UTs). To adapt our setup to the ATs, a custom mechanical interface was designed to direct starlight into a modified version of our fiber injection and spectral filtering coupling assemblies. For these observations the coupling assemblies were set up with H\,$\alpha$ filters with a bandpass of $\Delta \lambda \sim 1$\,nm. The fiber injection setup for both telescopes were  installed inside the Relay Optics Structure (ROS) of the ATs. The interface is easily removable such that the optical configuration of the ROS can be quickly reset to its original state. The interface consists primarily of a dichroic that is mounted in the beam path that redirects light onto our coupling assembly. A CAD drawing and photograph of the interface as installed in the ROS are pictured in Figure~\ref{fig:VLT_image}. 

\begin{figure}[t]
  \begin{subfigure}[t]{.5\textwidth}
    \centering
    \includegraphics[width=\linewidth]{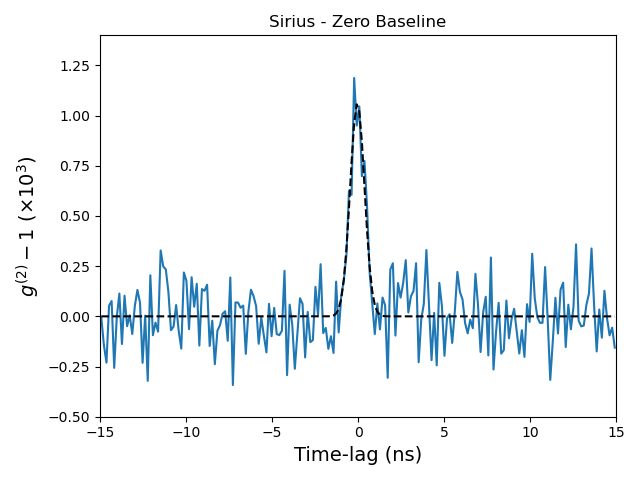}

  \end{subfigure}
  \hfill
  \begin{subfigure}[t]{.5\textwidth}
    \centering
    \includegraphics[width=\linewidth]{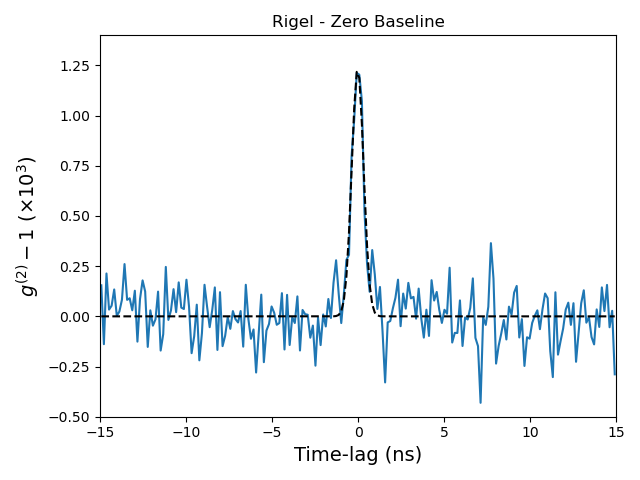}
  \end{subfigure}

  \begin{subfigure}[t]{.5\textwidth}
    \centering
    \includegraphics[width=\linewidth]{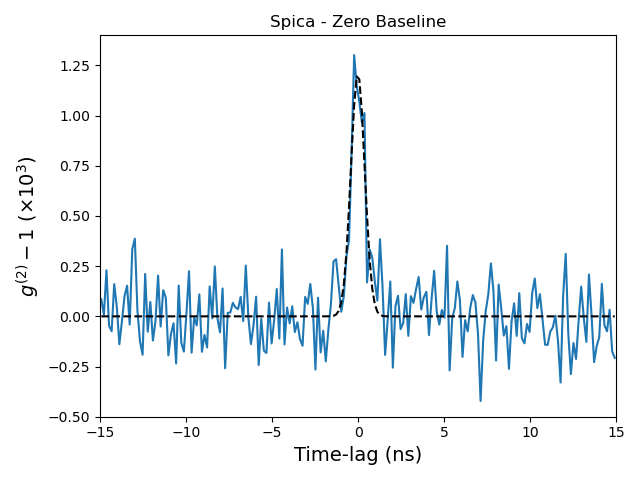}
  \end{subfigure}
  \hfill
  \begin{subfigure}[t]{.5\textwidth}
    \centering
    \includegraphics[width=\linewidth]{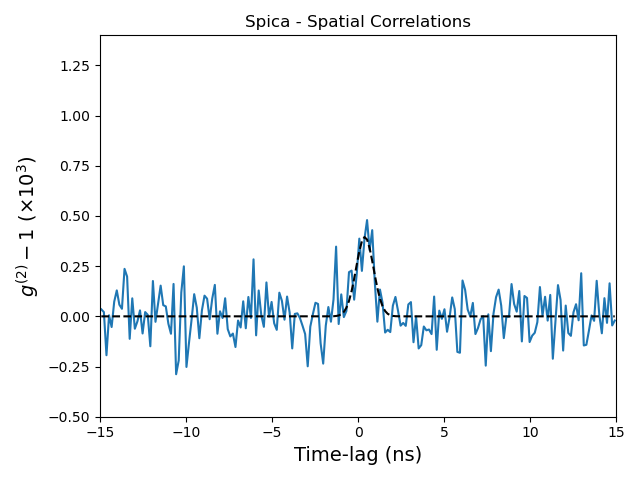}

  \end{subfigure}
\caption{\label{fig:VLT_data} Temporal and spatial $g^{(2)}$ measurements of three different stars using two of the VLTI-AT telescopes. The top-left, top-right, and bottom-left plots show the zero-baseline correlations measured for Sirius, Rigel, and Spica, each fit with a Gaussian function shown in the dotted black line. The bottom-right plot shows the averaged spatial correlation between the ATs of Spica that reveals a loss of contrast in the $g^{(2)}$ peak associated with a reduction in the visibility.}

\end{figure}

The entire system was installed during a single afternoon using an artificial light beacon to test and align the modules. The first night was used to ensure proper on-sky alignment and to assess the count rate on several sources. The following 5 nights were devoted to performing HBT measurements of stellar targets. In total we obtained 2.3 hours on Alf CMa (Sirius), 12.9 hours on Bet Ori (Rigel), and 34.3 hours on Alf Vir (Spica). For each source, we were able to resolve the zero-baseline $g^{(2)}(\tau)$ bunching peaks shown in Figure~\ref{fig:VLT_data}. To evaluate the bunching area we performed a Gaussian fit to the averaged correlation function obtained for each star. The bunching area measured was then compared against the expected value calculated from the spectrum. The spectrum was formed by combining the narrow-band filter transmission and target spectra that was contemporaneously measured with the interferometric observations. For the bunching areas we find for Sirius $\tau_m = 1.11 \pm 0.09\,$ps ($\tau_s = 1.14\,$ps), for Rigel $\tau_m = 1.07 \pm 0.07$\,ps ($\tau_s = 1.17\,$ps), and for Spica $\tau_m = 1.22 \pm 0.09$\,ps ($\tau_s = 1.12\,$ps), where $\tau_m$ is the measured area, and $\tau_s$ is the expected value from the spectrum. All measured values are in reasonable agreement with the expectation within 1.4$\,\sigma$. 

In addition, for both Rigel and Spica, $g^{(2)}(r)$ peaks for the correlations between the two ATs were measured, each showing a contrast less than for the zero-baseline calibration. For Rigel, a marginal peak was observed consistent with the low expected squared visibilities ($V^2 \sim$ 0.15) for the obtained projected baselines. In the case of Spica, a clear $g^{(2)}$ peak was observed with a significance of $7.5$\,$\sigma$ and shown in Figure~\ref{fig:VLT_data}. The bunching area corresponds to a measured squared visibility of $V^2 = 0.42\pm0.06$ that includes all data sampled over the baseline range from $\sim$\,23 to 48\,m. The expectation of the visibility of Spica is highly non-trivial as the system consists of a short period interacting binary of two hot B-type stars~\cite{HerbisonEvans1971,Aufdenberg2021} and a full analysis is beyond the scope of these proceedings.  

\section{Conclusion and Outlook}

The I2C system has been successfully tested with several of the facilities available on the Calern Plateau and additionally with two of the ATs of the VLTI. These results have demonstrated the versatility of the intensity interferometry technique to be utilized on telescopes of different optical designs. Several technical improvements are being pursued in order to improve sensitivity to allow for the observation of fainter systems or equivalently increase the precision of the squared visibility on stars of similar magnitude. Our current detectors exhibit a high quantum efficiency $> 60\%$ but with a fairly modest temporal resolution of $\sim 500$\,ps. In the near future we aim towards on-sky tests using super conducting nanowire single photon detectors which exhibit similar, if not better, quantum efficiency but with drastically better timing $< 50\,$ps\cite{Chang2019}. Furthermore, in order to have a sensitivity competitive with other HBT observatories using Cherenkov telescopes and ultimately with direct interferometry, it is necessary to implement spectral multiplexing. The sensitivity of an intensity interferometer is to first-order independent of the width of the spectral passband. Therefore, the sensitivity can be improved by performing $g^{(2)}$ measurements in many spectral passbands and co-adding the results~\cite{Trippe2014}. Some initial tests towards multiplexed observations are being explored in the laboratory using the spectrograph described earlier, as well as with comercially available monochromators. In addition, we will work towards scaling our system to work with 3 or 4 telescopes in order to provide improved simultaneous coverage of the uv-plane during observations.

\acknowledgments
We acknowledge the financial support of the Région PACA (project I2C), the French ANR (project I2C, ANR-20-CE31-0003), OCA, Doeblin federation and UCA science councils grants. The authors would like to thank Andreas Kaufer for the opportunity to test our equipment with the VLTI-AT telescopes. We acknowledge the use of spectral data obtained from the Southern Spectroscopic Project Observatory team\footnote{https://2spot.org/FR/}. This research has made use of the Jean-Marie Mariotti Center \texttt{Aspro} service\footnote{Available at http://www.jmmc.fr/aspro}. The authors would also like to thank Jason Aufdenberg for his assistance in our analysis of Spica data. 

\newpage
\bibliography{report} 

\begin{thebibliography}{10}

\bibitem{HBT1974}
Hanbury~Brown, R., Davis, J., and Allen, L.~R., ``The angular diameters of 32
  stars,'' {\em \mnras}~{\bf 167}(1),  121 (1974).

\bibitem{Labeyrie1975}
{Labeyrie}, A., ``{Interference fringes obtained on Vega with two optical
  telescopes.},'' {\em \apj}~{\bf 196},  L71--L75 (Mar 1975).

\bibitem{LeBohec2006}
LeBohec, S. and Holder, J., ``Optical intensity interferometry with atmospheric
  cerenkov telescope arrays,'' {\em \apj}~{\bf 649},  399–405 (Sep 2006).

\bibitem{Acciari2019}
{Acciari}, V.~A., {Bernardos}, M.~I., {Colombo}, E., {Contreras}, J.~L.,
  {Cortina}, J., {De Angelis}, A., {Delgado}, C., {D{\'\i}az}, C., {Fink}, D.,
  {Mariotti}, M., {Mangano}, S., {Mirzoyan}, R., {Polo}, M., {Schweizer}, T.,
  and {Will}, M., ``{Optical intensity interferometry observations using the
  MAGIC Imaging Atmospheric Cherenkov Telescopes},'' {\em \mnras}~{\bf 491},
  1540--1547 (Jan. 2020).

\bibitem{Abeysekara2020}
{Abeysekara}, A.~U., {Benbow}, W., {Brill}, A., {Buckley}, J.~H.,
  {Christiansen}, J.~L., {Chromey}, A.~J., {Daniel}, M.~K., {Davis}, J.,
  {Falcone}, A., {Feng}, Q., {Finley}, J.~P., {Fortson}, L., {Furniss}, A.,
  {Gent}, A., {Giuri}, C., {Gueta}, O., {Hanna}, D., {Hassan}, T., {Hervet},
  O., {Holder}, J., {Hughes}, G., {Humensky}, T.~B., {Kaaret}, P., {Kertzman},
  M., {Kieda}, D., {Krennrich}, F., {Kumar}, S., {LeBohec}, T., {Lin},
  T.~T.~Y., {Lundy}, M., {Maier}, G., {Matthews}, N., {Moriarty}, P.,
  {Mukherjee}, R., {Nievas-Rosillo}, M., {O'Brien}, S., {Ong}, R.~A., {Otte},
  A.~N., {Pfrang}, K., {Pohl}, M., {Prado}, R.~R., {Pueschel}, E., {Quinn}, J.,
  {Ragan}, K., {Reynolds}, P.~T., {Ribeiro}, D., {Richards}, G.~T., {Roache},
  E., {Ryan}, J.~L., {Santander}, M., {Sembroski}, G.~H., {Wakely}, S.~P.,
  {Weinstein}, A., {Wilcox}, P., {Williams}, D.~A., and {Williamson}, T.~J.,
  ``{Demonstration of stellar intensity interferometry with the four VERITAS
  telescopes},'' {\em \natast}~{\bf 4},  1164--1169 (July 2020).

\bibitem{Dravins2013}
Dravins, D., LeBohec, S., Jensen, H., and Nuñez, P.~D., ``Optical intensity
  interferometry with the cherenkov telescope array,'' {\em \app}~{\bf 43},
  331 -- 347 (2013).
\newblock Seeing the High-Energy Universe with the Cherenkov Telescope Array -
  The Science Explored with the CTA.

\bibitem{Guerin2017}
Guerin, W., Dussaux, A., Fouché, M., Labeyrie, G., Rivet, J.-P., Vernet, D.,
  Vakili, F., and Kaiser, R., ``Temporal intensity interferometry: photon
  bunching in three bright stars,'' {\em \mnras}~{\bf 472}(4),  4126--4132
  (2017).

\bibitem{Guerin2018}
Guerin, W., Rivet, J.-P., Fouché, M., Labeyrie, G., Vernet, D., Vakili, F.,
  and Kaiser, R., ``{Spatial intensity interferometry on three bright stars},''
  {\em \mnras}~{\bf 480},  245--250 (07 2018).

\bibitem{Rivet2020}
{Rivet}, J.~P., {Siciak}, A., {de Almeida}, E.~S.~G., {Vakili}, F., {Domiciano
  de Souza}, A., {Fouch{\'e}}, M., {Lai}, O., {Vernet}, D., {Kaiser}, R., and
  {Guerin}, W., ``{Intensity interferometry of P Cygni in the H
  {\ensuremath{\alpha}} emission line: towards distance calibration of LBV
  supergiant stars},'' {\em \mnras}~{\bf 494},  218--227 (Feb. 2020).

\bibitem{deAlmeida2022}
de~Almeida, E. S.~G., Hugbart, M., Domiciano~de Souza, A., Rivet, J.-P.,
  Vakili, F., Siciak, A., Labeyrie, G., Garde, O., Matthews, N., Lai, O.,
  Vernet, D., Kaiser, R., and Guerin, W., ``{Combined spectroscopy and
  intensity interferometry to determine the distances of the blue supergiants P
  Cygni and Rigel},'' {\em Monthly Notices of the Royal Astronomical Society}
  (June 2022).
\newblock https://doi.org/10.1093/mnras/stac1617.

\bibitem{Siegert1943}
Siegert, A. J.~F., ``On the fluctuations in signals returned by many
  independently moving scatterers, report / radiation laboratory, massachusetts
  institute of technology,'' (1943).

\bibitem{Ferreira2020}
Ferreira, D., Bachelard, R., Guerin, W., Kaiser, R., and Fouch\'{e}, M.,
  ``Connecting field and intensity correlations: The siegert relation and how
  to test it,'' {\em American Journal of Physics}~{\bf 88}(10),  831--837
  (2020).

\bibitem{Mandel1995}
Mandel, L. and Wolf, E.,  [{\em Optical Coherence and Quantum
  Optics}{\nolinebreak\hspace{0.1em}]}, Cambridge University Press (1995).

\bibitem{Bourges2016}
Bourg\`{e}s, L. and Duvert, G., ``{ASPRO2: get ready for VLTI's instruments
  GRAVITY and MATISSE},'' in [{\em Optical and Infrared Interferometry and
  Imaging V}{\nolinebreak\hspace{0.1em}]},  Malbet, F., Creech-Eakman, M.~J.,
  and Tuthill, P.~G., eds.,  {\bf 9907},  249 -- 259, International Society for
  Optics and Photonics, SPIE (2016).

\bibitem{HerbisonEvans1971}
Herbison-Evans, D., Hanbury~Brown, R., Davis, J., and Allen, L.~R., ``{A Study
  of $\alpha$ Virginis with an Intensity Interferometer},'' {\em \mnras}~{\bf
  151},  161--176 (01 1971).

\bibitem{Aufdenberg2021}
Aufdenberg, J.~P. and Hammill, J.~M., ``Modeling the h$\alpha$ emission
  surrounding spica using the lyman continuum from a gravity-darkened central
  star,'' {\em The Astrophysical Journal}~{\bf 923},  10 (dec 2021).

\bibitem{Chang2019}
Chang, J., Zadeh, I.~E., Los, J. W.~N., Zichi, J., Fognini, A., Gevers, M.,
  Dorenbos, S., Pereira, S.~F., Urbach, P., and Zwiller, V.,
  ``Multimode-fiber-coupled superconducting nanowire single-photon detectors
  with high detection efficiency and time resolution,'' {\em Appl. Opt.}~{\bf
  58},  9803--9807 (Dec 2019).

\bibitem{Trippe2014}
{Trippe}, S., {Kim}, J.-Y., {Lee}, B., {Choi}, C., {Oh}, J., {Lee}, T., {Yoon},
  S.-C., {Im}, M., and {Park}, Y.-s., ``{Optical multi-channel intensity
  interferometry - or: how to resolve O-stars in the Magellanic Clouds},'' {\em
  Journal of The Korean Astronomical Society}~{\bf {47}},  235--253 (2014).

\end{thebibliography}
\bibliographystyle{spiebib} 

\end{document}